\documentclass[aps,prc,twocolumn,groupedaddress,showpacs]{revtex4}
\usepackage{epsfig}

\begin{document}


\title{Relativistic model for nuclear matter and atomic nuclei
with momentum-dependent self-energies}


\author{S. Typel}
\affiliation{Gesellschaft f\"{u}r Schwerionenforschung mbH (GSI), Theorie,
Planckstra\ss{}e 1, D-64291 Darmstadt, Germany}


\date{\today}

\begin{abstract}
The Lagrangian density of standard relativistic mean-field  (RMF) models
with density-dependent meson-nucleon coupling vertices is modified
by introducing couplings of the meson fields to derivative nucleon
densities. As a consequence, the nucleon self energies, that describe
the effective in-medium interaction, become  
momentum dependent. In this approach
it is possible to increase the effective (Landau) mass of the nucleons, that is
related to the density of states at the Fermi energy, as compared to 
conventional relativistic models. At the same time
the relativistic effective (Dirac) mass is kept small
in order to obtain a realistic strength of the spin-orbit interaction.
Additionally, the empirical Schr\"{o}dinger-equivalent central optical 
potential from Dirac phenomenology is reasonably well described.
A parametrization of the model is obtained by a fit to properties of
doubly magic atomic nuclei. 
Results for symmetric nuclear matter, neutron matter
and finite nuclei are discussed.
\end{abstract}


\maketitle



\section{Introduction}

The relativistic description of nuclear matter and atomic nuclei
advanced considerably in recent years by improving 
the underlying density functional \cite{Wal74,Ser86,Rei89,Rin96}. 
Originally, these nuclear systems
were treated in the language of a relativistic quantum field
theory (Quantum Hadrodynamics) based on a relativistic Lagrangian 
density that contains nucleons and mesons as degrees of freedom.
In recent years, the viewpoint has changed and the approach is
considered as an effective field theory, see e.g.\ \cite{Rus97}.
The model can be seen as a special
variant of density functional theory \cite{Dre90} 
and a connection to the underlying
theory of Quantum Chromodynamics was established \cite{Fin04}.

In principle, it is possible to
describe the ground state of nuclei exactly 
due to the Hohenberg-Kohn theorem \cite{Hoh64}
as soon as the correct functional of the local baryon density is known.
Unfortunately, there is
no explicit scheme for constructing such a functional. Nuclear theory
is concerned to develop a sufficiently accurate
approximation guided by experience and the physical knowledge of the system.
Instead of using  functionals based on the density alone, one can
start from a functional that contains additional
densities, e.g.\ the kinetic energy density. This approach automatically
leads to the Kohn-Sham equations for the nucleons \cite{Koh65}, i.e.\ a
Schr\"{o}dinger-type equation in the non-relativistic case and
a Dirac-type equation in the relativistic case, respectively.
In these equations the effective interaction is expressed by
potentials or by self-energies, respectively, that depend non-trivially
on the nucleon fields.

There is a big advantage of the Kohn-Sham approach since certain physical
effects are more easily incorporated into the description.
Functionals that are solely based on the density can be compared to
a description of finite nuclei in a Thomas-Fermi approximation
where no shell effects appear. Considering also the kinetic density in the
functional, corresponding to a Hartree or Hartree-Fock description,
shell effects emerge automatically.
In non-relativistic approaches a spin current density is included
in the functional to generate the large spin-orbit splittings
observed in finite nuclei. These three densities in the non-relativistic
framework are replaced by the baryon (or vector) density, 
the kinetic density and
the scalar density in relativistic approaches.
Pairing effects can be included by introducing the corresponding
pairing densities.
Additionally, in case of neutron-proton asymmetric systems,
one has to deal
with both isoscalar and isovector versions of these quantities.

A general functional can be constructed in terms of the
considered densities and their derivatives.
Due to the Lorentz structure and the isospin degree of freedom there
is a large variety of possible contributions that can appear in a
relativistic density functional and it is not clear what possible
densities and combinations are really relevant in the description.
The density functional can be expanded in powers of the nucleon field, 
the meson fields, and their derivatives. The importance of the individual
contributions can be estimated
in a systematic approach that is guided by the principles
of naive dimensional analysis and of the naturalness 
of the appearing coupling constants \cite{Rus97,Fur97,Fri96}.
It is also possible to compare with more fundamental approaches
based on QCD and chiral symmetry.
For large densities, however, this approach might be problematic
due to the resulting polynomial dependence of the functional on the
Fermi momentum or the density.
Well-behaved rational approximations seem to be a promising alternative.
Extrapolations to high densities are necessary in the application of the model,
e.g.\ in the description of heavy-ion collisions or neutron stars.

The usual starting point in relativistic models is a certain Lagrangian
density instead of an equivalent energy functional. This Lagrangian
can be treated in different approximations. For a practical application
to nuclear matter and finite nuclei one usually employs the mean-field
approximation. In this approach the nucleon self-energies in nucler matter
become simple functions of various nucleon densities. In principle, one can
go beyond the mean-field approximation in a quantum field theoretical
treatment by a systematic diagrammatic expansion. In this case, the
nucleon self-energies will have a much more complicated structure that
takes the non-trivial effects of the nuclear medium into account.
In Dirac Brueckner calculations of nuclear matter based on a given
nucleon-nucleon interaction it is possible to extract
the corresponding nucleon self energies, see, e.g., 
\cite{Seh97,deJ98,Gro99,Hof01,Dal04}. 
They show a dependence
on both the density of the nuclear medium and the energy and momentum of the
nucleon. An alternative approach to these ``ab-initio'' 
strategies for the nuclear 
many-body problem is a phenomenological treatment of the problem
where the mean-field approximation is retained with its virtues in applying
the model self-consistently to nuclear matter and finite nuclei with
reasonable effort. However, the basic Lagrangian has to be modified for
a quantitative description of the various properties and it becomes
mandatory to introduce new terms or to modify existing contributions
in the Lagrangian. The parameters
of the model have to be fitted to properties of nuclear matter and
atomic nuclei. They cannot be derived in a simple manner
from a more basic description. In this work the phenomenological 
strategy is followed since it is known from experience that
this approach is very successful.
A selection of the relevant terms and of a more generalized functional
form  that is guided by other approaches, e.g.\ Dirac Brueckner 
calculations, and by physical intuition may lead to 
reasonable results. 

In relativistic approaches to nuclear structure it is common to
introduce meson fields as explicit degrees of freedom in the
Lagrangian or density functional. In the modern point of view, they have to be 
seen as auxiliary fields 
that, a priori, have not to be identified with the corresponding 
mesons in free space although they share the same Lorentz and isospin 
structure. In principle, these fields can be completely removed from the 
theory as, e.g., in the relativistic point-coupling models
\cite{Nik92,Fri96,Rus97,Bue02}.
In the following, however, we will keep these meson fields in the model
for convenience.

A quantitative description of nuclear systems was achieved by
considering an explicit medium dependence of the effective interaction.
For that purpose non-linear self-interactions of the 
mesons were introduced and the corresponding 
coupling constants were fitted to properties of nuclear matter 
and finite nuclei mostly close to the valley of stability, see e.g.\ 
\cite{Bog77,Bog81,Bog83,Rei86,Ruf88,Rei88,Bod89,Bod91,Gmu91,Sha92,Sha93,%
Sug94,Kou95,Lal97,Tok98,Lon04}. At large
densities, however, this approach is not really reliable due to the
polynomial dependence on the fields and instabilities can occur.
Alternatively, a density dependence of the meson-nucleon couplings 
was considered 
\cite{Mar89,Seh90,Bro92,Had93,Fri94,Fuc95,Typ99,Hof01,Nik02,Lon04,Lal05}.
In Ref.\ \cite{Fuc95} it has been pointed out that,
in general, the couplings have to be treated as functionals of the
nucleon fields leading to rearrangement contributions to the self-energies
that are necessary for the thermodynamical consistency of the model.
It is possible to choose functional forms (e.g.\ well-behaved rational
functions) that are motivated by results of Dirac-Brueckner calculations 
of nuclear matter \cite{Typ99,Hof01}.
This class of models seems to be the more flexible approach
with proper high density behavior. Originally, the medium dependence 
was introduced only in the isoscalar part of the interaction.
In recent years it was realized that it is also necessary
to allow for a density dependence in the isovector channel in order
to obtain a reasonable description of the neutron
skin thickness of stable nuclei, the neutron matter equation of state
and a reliable extrapolation to exotic nuclei \cite{Typ01,Fur02}.
In all these standard relativistic approaches
the self-energies in the Dirac equation for the nucleons depend
nontrivially on the various densities. However, they are the same
for all protons and neutrons, respectively, independent of the
single-particle state.

Despite the success of the relativistic approach
there are still some deficiencies 
that have to be dealt with in further extensions of the underlying
density functional. The size of the 
scalar ($\Sigma$) and vector ($\Sigma_{\mu}$)
self-energies in the interior of a finite nucleus is well determined
by requiring that the corresponding central and spin-orbit potentials
have the correct strength.
(For alternative definitions of the non-relativistic potential
see, e.g., Ref.\ \cite{Typ02}.)
They are derived in a nonrelativistic reduction of the Dirac equation.
The Schr\"{o}dinger-equivalent
central potential
\begin{equation} \label{eq:pot_cen}
 V_{\rm cen} = \frac{E}{m} \Sigma_{0} - \Sigma 
 + \frac{1}{2m} \left( \Sigma^{2} - \Sigma_{0}^{2}\right)
\end{equation}
and the spin-orbit potential
\begin{equation}
 V_{\rm so} = \frac{\frac{d}{dr}\left(\Sigma_{0}+\Sigma\right)}{
 E + m - \Sigma_{0} - \Sigma} \frac{\vec{\sigma} \cdot \vec{L}}{2mr}
\end{equation}
of a nucleon with restmass $m$ and energy $E$ in a spherical nucleus
are determined by the sum and the difference of the time component
of the vector self-energy $\Sigma_{0}$ and the scalar self-energy $\Sigma$,
fixing both quantities.
Correspondingly, in symmetric nuclear matter at saturation density 
$\varrho_{\rm sat}=2k_{F}^{3}/(3\pi^{2})$ with Fermi momentum
$p_{F} = \hbar k_{F}$ 
the relativistic effective (Dirac) mass 
\begin{equation} \label{eq:rem}
 m^{\ast} = m - \Sigma
\end{equation}
that is independent of the nucleon momentum
and the chemical potential
\begin{equation}
 \mu = \Sigma_{0} + \sqrt{\left(m^{\ast}\right)^{2}+p_{F}^{2}}
\end{equation}
are well determined. 

In conventional RMF models the self-energies
are independent of the nucleon energy and momentum but they
exhibit a strong density dependence.
However, from Dirac-Brueckner calculations one would expect that
the self-energies also depend on the energy and momentum of the nucleon
in the medium. The central potential (\ref{eq:pot_cen}) allows to
quantify this effect in both relativistic and non-relativistic models.
Standard RMF models show a linear increase of the
central potential. Only models with explicit energy or momentum
dependent self-energies show a different result.
In order to compare the energy or momentum dependence of (\ref{eq:pot_cen})
with empirical data it is common practice to compare the model predictions
with the optical potential in nuclear matter extracted from Dirac
phenomenology (DP) for elastic proton-nucleus scattering \cite{Ham90,Coo93}.
In this approach scattering observables are 
fitted to experimental data up to approx.\ 1~GeV kinetic energy by varying
the strength of the vector and scalar self-energies (real and imaginary
part) of the target. The potential shape is described in
simple parametrizations \cite{Ham90,Coo93}
or taken from microscopic descriptions \cite{Typ02}. 
In DP the self-energies depend on the proton energy. The derived
Schr\"{o}dinger-equivalent central potential, extrapolated
to nuclear matter, exhibits a non-linear energy dependence.
At high energies a saturation of the optical potential 
with a value around 50~MeV
is observed. (Imaginary contributions of the self-energies 
as extracted from Dirac phenomenology
have an almost
negligible influence on the real part of the corresponding optical potential.
They enter only through the terms quadratic in the self-energies
in Eq.\ (\ref{eq:pot_cen}). These contributions almost cancel each other
since they are of similar magnitude.)
At low kinetic energies, the optical potential is a nearly linear function
of the energy. The self-energies in DP itself show an almost linear
energy dependence for not too high energies.
This observation and the experience from Dirac-Brueckner calculations
motivates to consider a phenomenological extension of
standard relativistic models that generates energy or
momentum dependent self-energies already on the mean-field level.

Besides the relativistic effective (Dirac) mass $m^{\ast}$ there are a number
of other effective masses that have been introduced in the literature.
For a general overview of definitions for effective
masses in relativistic and non-relativistic calculations see
\cite{Jam89}. An important quantity is
the effective Landau mass 
\begin{equation}
 m_{\rm eff} = p \frac{dp}{dE} 
\end{equation}
that is related to the density of states 
$\varrho = m_{\rm eff} p/(2\pi\hbar)^{3}$
in nuclear matter.
The Landau mass is easily calculated from the dispersion
relation
\begin{equation}
 \left(E-\Sigma_{0}\right)^{2} = \left( m - \Sigma \right)^{2}
 + p^{2}
\end{equation}
for a nucleon. In relativistic models without energy or momentum
dependent self-energies one obtains
\begin{equation} \label{eq:meff2}
 m_{\rm eff} = E-\Sigma_{0}
 = \sqrt{\left(m^{\ast}\right)^{2} + p_{F}^{2}} 
\end{equation}
at the Fermi momentum $p_{F}$.
The relativistic effective mass (\ref{eq:rem}) cannot be adjusted
arbitrarily if a reasonable description of the spin-orbit interaction
in atomic nuclei is required. 
In usual relativistic models we find $m^{\ast} \approx 0.55 m$ and
$m_{\rm eff} \approx 0.62 m > m^{\ast}$ at the Fermi momentum
but the latter value is rather small.
Comparing non-relativistic Skyrme Hartree-Fock calculations
for giant resonances in the random phase approximation
with experimental data 
a value $0.78m$ for $m_{\rm eff}$ was extracted \cite{Rei99}.
The low Landau mass in conventional RMF models
indicates that the level density at the Fermi energy 
is too small and single-particle levels 
in finite nuclei are too much spreaded, a well known
result of detailed calculations.

Assuming an energy dependence of the self-energies,
the effective Landau mass is given by
\begin{equation} \label{eq:meff}
 m_{\rm eff} = 
  (E-\Sigma_{0})\left(1-\frac{d\Sigma_{0}}{dE}\right)
 + (m-\Sigma)\frac{d\Sigma}{dE} \: .
\end{equation}
It is clear that the Landau mass can be adjusted more freely
if the model allows for  energy dependent self energies.
However, one can not expect that the density and energy/momentum
dependence of the Landau mass as observed, e.g., in DB calculations
can be well represented.
The result (\ref{eq:meff}) is consistent with the non-relativistic form
\begin{equation} 
 m_{\rm eff} =
 m \left( 1 - \frac{dV}{dE} \right)
\end{equation}
with an energy-dependent optical potential $V$ taken from (\ref{eq:pot_cen})
in the non-relativistic approximation.

There are attempts to cure the problem of the small Landau mass
and the related low density of levels
in relativistic models.
In Ref.\ \cite{Vre02} a linear energy dependence of the self-energies
was introduced heuristically in an energy window around the Fermi energy
in order to increase the effective (Landau) mass and the level density.
However, the full self-consistency  and the Lorentz invariance 
of the RMF model are lost. In another approach 
the Landau mass $m_{\rm eff}$ was increased to the value
of $0.76 m$ that was necessary to obtain
reasonable $\beta$ decay half-lives \cite{Nik04}.
Correspondingly, the  Dirac effective mass 
had to be increased by hand to $m^{\ast}= 0.67 m$ destroying the 
usually good description of the spin-orbit splittings in nuclei
without additional modifications of the density functional.
The authors were able to compensate 
this reduction of the scalar self-energy 
by introducing a rather strong tensor interaction in the model
but still retaining a reasonable description of other properties.

Considering the above observations it is natural to extend the 
standard relativistic density functional in a way that 
energy or momentum dependent self-energies appear in the Kohn-Sham equations.
This will be a phenomenological approach introducing new parameters
in the model that have to be fitted to properties of nuclear matter 
and atomic nuclei.
In Ref.\ \cite{Typ03} couplings of the meson fields to derivatives
of the nucleon field were introduced in the Lagrangian density
with the desired result. In this derivative coupling (DC) model
additional densities in the functional appear that were not
considered before. It was possible to describe the experimentally
observed energy dependence of the optical potential
by adjusting the relevant coupling constants but still retaining reasonable
nuclear matter properties. There are, however, some 
deficiencies of the approach. The density dependence of the
momentum dependence was different for scalar and vector self-energies
leading to problems in extrapolations of the model to high densities.
Since the relativistic energy  of the nucleon contains the rest mass
rather large contributions of the energy-dependent part
of the self-energy are required to show a sizable effect. Simultaneously,
a large modification of the standard energy-independent part of
the self-energies was required. An attempt to fit the parameters of the model
to properties of finite nuclei proved to be very difficult.

Realizing the problems, this paper introduces a modification of the
model in Ref.\ \cite{Typ03} combining the virtues of the 
density-dependent approach for the coupling constants with the
momentum-dependent self-energies. 
In Sec.\ \ref{sec:1} the basic
Lagrangian density of the model with density-dependent
and derivative couplings (D${}^{3}$C) 
is introduced and the field equations for nucleons,
mesons and the photon are derived. The relevant equations
for nucler matter and finite nuclei are presented in Sec.\ \ref{sec:2}.
The parametrization of the coupling functions and the fit of the
parameters to properties of finite nuclei are described in Sec.\ \ref{sec:3}.
Results of the model for nuclear matter and finite nuclei are 
discussed in Sec.\ \ref{sec:4} with a comparision to conventional RMF models.
Finally, in Sec.\ \ref{sec:5}, conclusions and an outlook complete
the paper.

\section{\label{sec:1} Lagrangian density and field equations}

In the present approach the (symmetrized) Lagrangian density assumes the form
\begin{equation} \label{eq:Lag}
 \mathcal{L} = 
 \frac{1}{2} \left[  \bar{\psi}  \Gamma_{\mu} i D^{\mu} \psi
 + \overline{(i D^{\mu}\psi)}  \Gamma_{\mu}  \psi \right]
 -  \bar{\psi} \Gamma M^{\ast} 
 \psi + {\cal L}_{m}
\end{equation}
with the nucleon field $\psi$, the covariant derivative 
\begin{equation}
 iD_{\mu} = 
 i \partial_{\mu} -\Gamma_{\omega} \omega_{\mu}
 - \Gamma_{\rho}\vec{\tau} \cdot \vec{\rho}_{\mu} 
 - \Gamma_{\gamma} \frac{1+\tau_{3}}{2} A_{\mu} 
\end{equation}
and the mass operator
\begin{equation}
 M^{\ast} = m-\Gamma_{\sigma}\sigma
  -\Gamma_{\delta} \vec{\tau} \cdot \vec{\delta} \: .
\end{equation}
The meson fields are denoted by $\sigma$, $\omega_{\mu}$,
$\vec{\delta}$,
$\vec{\rho}_{\mu}$ and the photon field by $A_{\mu}$.
For completeness all four combinations from the alternatives
scalar-vector and isoscalar-isovector are included for the mesons.
The quantities $\Gamma_{\sigma}$, $\Gamma_{\omega}$,
$\Gamma_{\delta}$, $\Gamma_{\rho}$ and $\Gamma_{\gamma}$
specify the coupling strenght of the mesons and of the photon, respectively,
to the nucleon. The elements of the vector $\vec{\tau}$ are
the isospin matrices.
The  contribution
\begin{eqnarray}
 \mathcal{L}_{m} & = & \frac{1}{2} 
 \left[ \partial^{\mu} \sigma \partial_{\mu} \sigma
  - m_{\sigma}^{2} \sigma^{2} 
+\partial^{\mu} \vec{\delta} \cdot \partial_{\mu} \vec{\delta}
  - m_{\delta}^{2} \vec{\delta} \cdot \vec{\delta}
  \right. \\ \nonumber & & 
- \frac{1}{2} F^{\mu \nu} F_{\mu \nu} 
  - \frac{1}{2} G^{\mu \nu} G_{\mu \nu} 
  + m_{\omega}^{2} \omega^{\mu} \omega_{\mu} 
 \\ \nonumber & & \left.
  - \frac{1}{2} \vec{H}^{\mu \nu} \cdot \vec{H}_{\mu \nu} 
  + m_{\rho}^{2} \vec{\rho}^{\mu} \cdot \vec{\rho}_{\mu} 
 \right]
\end{eqnarray} 
is the Lagrangian density for the free mesons with
masses $m_{\sigma}$, $m_{\omega}$, $m_{\delta}$, $m_{\rho}$, and the photon
with the field tensors 
\begin{eqnarray}
   F_{\mu \nu}  & = &
  \partial_{\mu} A_{\nu} - \partial_{\nu} A_{\mu} 
 \\ 
  G_{\mu \nu} & = &
  \partial_{\mu} \omega_{\nu} - \partial_{\nu} \omega_{\mu} 
 \\
  \vec{H}_{\mu \nu} & = &
  \partial_{\mu} \vec{\rho}_{\nu} - \partial_{\nu} \vec{\rho}_{\mu} \: .
\end{eqnarray}

In standard relativistic models the quantities $\Gamma_{\mu}$ and
$\Gamma$ in Eq.\ (\ref{eq:Lag})
are the Dirac matrices $\gamma_{\mu}$ and the unit matrix, respectively.
In the present approach, they are given by
\begin{eqnarray} \label{eq:gam_a}
 \Gamma_{\mu} & = & 
  \gamma^{\nu} g_{\mu \nu} + \gamma^{\nu} Y_{\mu \nu} 
 - g_{\mu \nu} Z^{\nu} 
 \\ \label{eq:gam_b}
 \Gamma & = & 
 1 + \gamma_{\mu} u_{\nu} Y^{\mu \nu}  
 - u_{\mu} Z^{\mu} 
\end{eqnarray}
with the quantities
\begin{eqnarray}
 Y^{\mu \nu} & = & \frac{\Gamma_{V}}{m^{4}} 
 m_{\omega}^{2} \omega^{\mu} \omega^{\nu}
 \\
 Z^{\mu} & = & \frac{\Gamma_{S}}{m^{2}}  \omega^{\mu} \sigma
\end{eqnarray}
that depend on the isoscalar $\sigma$ and $\omega$ meson fields.
An extension of the approach to isovector fields is obvious.
The particular dependence on the meson fields is required 
by the Lorentz structure
in order to generate a linear energy dependence of the scalar and
vector self-energies (see below). Furthermore, for both $Y^{\mu \nu}$
and $Z^{\mu}$ a form quadratic in the meson fields was chosen
so that the density dependence
of the energy dependence is similar for both self-energies
in contrast to the DC model \cite{Typ03}.
$\Gamma_{V}$ and $\Gamma_{S}$ represent the two additional
couplings of the D${}^{3}$C model that can depend on
the nucleon fields similar as the quantities 
$\Gamma_{\sigma}$, $\Gamma_{\omega}$, $\Gamma_{\delta}$, and
$\Gamma_{\rho}$ in the minimal coupling of the nucleon field
to the meson fields.
In Eqs.\ (\ref{eq:gam_a}) and (\ref{eq:gam_b}) 
the metric tensor $g_{\mu \nu} = \mbox{diag} (1,-1,-1,-1)$
and the four velocity $u_{\mu} = j_{\mu}/\varrho_{v}$ depending on
the vector current density $j_{\mu} = \bar{\psi} \gamma_{\mu} \psi$
with the vector density $\varrho_{v} = \sqrt{j_{\mu}j^{\mu}}$
appear. Combining the usual minimal meson-nucleon couplings with the
action of the quantities $\Gamma_{\mu}$ and $\Gamma$ one finds
that terms with third powers in the meson fields appear in 
the Lagrangian like in the non-linear models with meson self interactions.
But in the present approach there are no free parameters for these individual
terms and they are all accompanied by nuclear fields. 
Similarly to the standard relativistic models
with density-dependent couplings we assume that the couplings 
$\Gamma_{\sigma}$, $\Gamma_{\omega}$,
$\Gamma_{\delta}$, and $\Gamma_{\rho}$ as well as
$\Gamma_{V}$ and $\Gamma_{S}$
depend on the vector density $\varrho_{v}$.
In principle, one can also imagine a dependence on other densities,
e.g., the scalar density $\varrho_{s} =  \bar{\psi} \psi$.
The explicit choice of the functional form will be discussed in Sec.\
\ref{sec:3}.

The Dirac equation, i.e.\ the Kohn-Sham equation, 
for the nucleons in the D${}^{3}$C model
\begin{equation} \label{eq:Dirac}
   \gamma^{\mu} \left( i \partial_{\mu}- \Sigma_{\mu} \right) \psi 
 - \left( m - \Sigma \right) \psi = 0
\end{equation}
has the standard form of relativistic models with the vector self-energy
\begin{equation} \label{eq:se_v}
 \Sigma_{\mu}  =    v_{\mu} 
   - Y_{\mu \nu} \left( i D^{\nu} 
 - M^{\ast}  u^{\nu}\right) 
 + \Sigma^{R}_{\mu}
\end{equation}
and the scalar self-energy
\begin{equation}
 \Sigma  =  s
 -  Z^{\mu} \left(  i D_{\mu} 
 - M^{\ast} u_{\mu}\right)  
\end{equation}
where
\begin{eqnarray} 
 v_{\mu} & = &   \Gamma_{\omega} \omega_{\mu}
 + \Gamma_{\rho}\vec{\tau} \cdot \vec{\rho}_{\mu} 
 + \Gamma_{\gamma} \frac{1+\tau_{3}}{2} A_{\mu}
 \\ \nonumber & & 
 - \frac{i}{2}  \partial^{\lambda} Y_{\mu \lambda}
\end{eqnarray}
and 
\begin{equation}
 s  =  \Gamma_{\sigma}\sigma
  +\Gamma_{\delta} \vec{\tau} \cdot \vec{\delta}
 - \frac{i}{2}   \partial^{\mu} Z_{\mu} \: .
\end{equation}
The self-energies in the D${}^{3}$C model
are differential operators that act on the nucleon field $\psi$.
The momentum dependence
enters through the contributions proportional to $\left(  i D_{\mu} 
 - M^{\ast} u_{\mu}\right)$. The time component of this expression
mimics the kinetic energy 
$E-m$ when applied to a plane wave in the absence of meson fields.
This is in contrast to the DC model where only the $i D_{\mu}$ term
appears \cite{Typ03}.

The vector self-energy (\ref{eq:se_v})
contains the rearrangement contribution
\begin{eqnarray} \label{eq:rearr}
 \lefteqn{\Sigma^{R}_{\lambda}   =  
   u_{\lambda} \left[  \Gamma^{\prime}_{\omega} \omega_{\mu} J^{\mu}
 + \Gamma^{\prime}_{\rho}\vec{\rho}_{\mu} \cdot \vec{J}^{\mu} 
 -  \Gamma_{\sigma}^{\prime} \sigma P_{s}
 - \Gamma^{\prime} _{\delta} \vec{\delta} \cdot \vec{P}_{s} \right.}
 \\ \nonumber & & \left.
 -  \left( t^{D}_{\nu\mu} - u_{\nu} j^{M^{\ast}}_{\mu} \right)
 \frac{\Gamma^{\prime}_{V}}{\Gamma_{V}} Y^{\mu \nu} 
 + \left( j^{D}_{\mu} - u_{\mu} \varrho^{M^{\ast}}_{s} \right)
 \frac{\Gamma^{\prime}_{S}}{\Gamma_{S}} Z^{\mu}
   \right] 
 \\ \nonumber & & 
 + \left( j^{M^{\ast}}_{\mu} Y^{\mu \nu}  
 - \varrho^{M^{\ast}}_{s} Z^{\nu} \right)
  \frac{g_{\nu \lambda}-u_{\nu}u_{\lambda}}{\varrho_{v}}
\end{eqnarray}
with derivatives $\Gamma_{i}^{\prime} = d\Gamma_{i}/d\varrho_{v}$
($i=\sigma$, $\omega$, $\delta$, $\rho$, $V$, $S$) and
densities
\begin{eqnarray} \label{eq:J}
 J_{\mu}  =   \bar{\psi} \Gamma_{\mu} \psi 
 & \qquad &
 \vec{J}_{\mu}  =   \bar{\psi} \Gamma_{\mu} \vec{\tau} \psi
 \\ \label{eq:P}
 P_{s}  =  \bar{\psi} \Gamma \psi 
 & \qquad &
 \vec{P}_{s}  =   \bar{\psi} \Gamma \vec{\tau} \psi 
\end{eqnarray}
that replace the usual vector and scalar densities.
Additionally, the densities
\begin{eqnarray} \label{eq:tD}
 t^{D}_{\mu \nu} & = & 
 \frac{1}{2} \left[
 \bar{\psi} \gamma_{\mu} i D_{\nu} \psi 
 + \overline{\left(  i D_{\nu} \psi \right)} \gamma_{\mu} \psi  \right]
 \\ \label{eq:jD}
 j^{D}_{\mu} & = & 
 \frac{1}{2} \left[ \bar{\psi} i D_{\mu} \psi 
 + \overline{\left(  i D_{\mu} \psi \right)} \psi  \right]
 \\ 
 j^{M^{\ast}}_{\mu} & = & 
 \bar {\psi} \gamma_{\mu} M^{\ast} \psi 
 \\
 \varrho^{M^{\ast}}_{s} & = & \bar {\psi} M^{\ast} \psi
\end{eqnarray}
appear in
the rearrangement contribution (\ref{eq:rearr}). It  
is slightly more complicated than in the standard
density-dependent (DD) models due to the occurence of the terms with 
$Y_{\mu \nu}$ and $Z_{\mu}$. It simplifies considerably for stationary
systems.
The kinetic densities (\ref{eq:tD}) and (\ref{eq:jD}) are different from
the derivatives $iD_{\nu} j_{\mu}$ and $iD_{\mu} \varrho_{s}$ 
of the vector current $j_{\mu}$ and scalar density
$\varrho_{s}$. The former do not
vanish in homogeneous nuclear matter and give a finite contribution.

From the Dirac equation (\ref{eq:Dirac}) the continuity equations
\begin{equation}
 \partial^{\mu} J_{\mu} = 0 
 \quad \mbox{and} \quad
 \partial^{\mu} \vec{J}_{\mu} = 0
\end{equation}
are derived. Correspondingly, $J_{0}$ and $\vec{J}_{0}$ are the
conserved isoscalar and isovector baryon densities instead of
$j_{0}=\varrho_{v}$ and $\vec{\jmath}_{0}= \bar{\psi} \gamma_{0}
\vec{\tau} \psi$ in standard RMF models.

In the D${}^{3}$C model
the field equations of the mesons are obtained as
\begin{eqnarray}
 \partial_{\mu} \partial^{\mu} \sigma +  m_{\sigma}^{2}  \sigma
 + \tilde{C}_{\mu} \omega^{\mu}  
 & = &   \Gamma_{\sigma} P_{s}
 \\ 
 \partial^{\nu} G_{\nu \mu}
 +  m_{\omega}^{2}
 \omega^{\nu}  C_{\mu \nu}  
 - \tilde{C}_{\mu}  \sigma
 & = &  \Gamma_{\omega} J_{\mu} 
 \\
  \partial_{\mu} \partial^{\mu} \vec{\delta} +  m_{\delta}^{2} \vec{\delta}
 & = &   \Gamma_{\delta} \vec{P}_{s} 
 \\ 
 \partial^{\nu} \vec{H}_{\nu \mu}
 +  m_{\rho}^{2} \vec{\varrho}_{\mu}  & = & 
 \Gamma_{\rho} \vec{J}_{\mu} 
\end{eqnarray}
with
\begin{equation}
 C_{\mu \nu} = g_{\mu \nu} + \frac{\Gamma_{V}}{m^{4}}
 \left( t^{D}_{\mu \nu} + t^{D}_{\nu \mu} 
 -  u_{\nu} j^{M^{\ast}}_{\mu}  -  u_{\mu} j^{M^{\ast}}_{\nu}  \right)
\end{equation}
and
\begin{equation}
\tilde{C}_{\mu} = \frac{\Gamma_{S}}{m^{2}}
 \left( j^{D}_{\mu}  - u_{\mu} \varrho^{M^{\ast}}_{s} \right)
 \: .
\end{equation}
The source terms are given by simple products of
the density-dependent coupling functions
$\Gamma_{\sigma}$, $\Gamma_{\omega}$, $\Gamma_{\delta}$, $\Gamma_{\rho}$,
and the densities (\ref{eq:J}) and (\ref{eq:P}).
For $\Gamma_{S} \neq 0$ there is a coupling of the equations for
the $\sigma$ and $\omega$ field.
Explicit terms
with $\Gamma_{V}$ appear only in the equation for the $\omega$ field due
to the factor  $C_{\mu \nu}$.
The field equation of the photon 
\begin{eqnarray}
 \partial^{\nu} F_{\nu \mu}
   & = & 
 \Gamma_{\gamma} J_{\gamma \mu} 
\end{eqnarray}
has the usual form with the conserved charge current density
$J_{\gamma \mu} = [J_{\mu}+(\vec{J}_{\mu})_{3}]/2$.

\section{\label{sec:2} Stationary systems}

The equations of motion for the nucleons,
mesons and the photon field simplify considerably if the nuclear system 
possesses certain symmetries. E.g., in stationary systems the meson fields
are independent of time. For nuclear matter and finite nuclei,
it suffices to consider only the time-like component of all
four-vectors and the third component of the isospin vectors.
The conserved baryon density
\begin{equation}
 \varrho = J_{0} = j_{0} \left( 1 + Y_{00} \right) - \varrho_{s} Z_{0}
\end{equation}
depends on the usual vector density $j_{0} = \varrho_{v}$ and the standard
scalar density $\varrho_{s}$.
Similarly, the generalized scalar density is given by the combination
\begin{equation}
 P_{s} = \varrho_{s} \left( 1 - Z_{0} \right) + j_{0} Y_{00} 
\end{equation}
with the quantities
\begin{equation}
 Y_{00}  =  \frac{\Gamma_{V}}{m^{4}} 
 m_{\omega}^{2} \omega_{0}^{2}
\end{equation}
and
\begin{equation}
 Z_{0}  =  \frac{\Gamma_{S}}{m^{2}}  \omega_{0} \sigma \: .
\end{equation}
Corresponding equations hold for the isovector densities
$\vec{\varrho}=\vec{J}_{0}$ and $\vec{P}_{s}$.
The self-energies in the Dirac equation can be written as
\begin{equation} \label{eq:sig0}
 \Sigma_{0}  =  V_{0} - Y_{00} i \partial^{0}  
\end{equation}
and 
\begin{equation} \label{eq:sig}
 \Sigma  =  S - Z_{0} i \partial^{0}
\end{equation}
with
\begin{equation} 
 V_{0}  =  
 v_{0} + Y_{00} \left(v_{0}+m-s\right)
 + \Sigma^{R}_{0}
\end{equation}
and 
\begin{equation} 
 S  =  
 s + Z_{0} \left(v_{0}+m-s\right)
\end{equation}
where
\begin{equation} 
 v_{0}  =    \Gamma_{\omega} \omega_{0}
 + \Gamma_{\rho}\vec{\tau} \cdot \vec{\rho}_{0} 
 + \Gamma_{\gamma} \frac{1+\tau_{3}}{2} A_{0}
\end{equation}
and 
\begin{equation}
 s  =  \Gamma_{\sigma}\sigma
  +\Gamma_{\delta} \vec{\tau} \cdot \vec{\delta}
\end{equation}
with the rearrangement contribution
\begin{eqnarray}
 \lefteqn{\Sigma^{R}_{0}   =  
   \left[  \Gamma^{\prime}_{\omega} \omega_{0} J^{0}
 + \Gamma^{\prime}_{\rho}\vec{\rho}_{0} \cdot \vec{J}^{0} 
 -  \Gamma_{\sigma}^{\prime} \sigma P_{s}
 - \Gamma^{\prime} _{\delta} \vec{\delta} \cdot \vec{P}_{s} \right.}
 \\ \nonumber & & \left.
 -  \left( t^{D}_{00} -  j^{M^{\ast}}_{0} \right)
 \frac{\Gamma^{\prime}_{V}}{\Gamma_{V}} Y_{00} 
 + \left( j^{D}_{0} -  \varrho^{M^{\ast}}_{s} \right)
 \frac{\Gamma^{\prime}_{S}}{\Gamma_{S}} Z_{0}
   \right]  \: .
\end{eqnarray}
The partial derivative $i \partial^{0}$ in (\ref{eq:sig0}) and (\ref{eq:sig}) 
gives the single-particle
energy $E$ when applied to a single-particle state in nuclear matter
or finite nuclei leading to an explicit energy dependence of the self-energies.
Note that the potentials $v_{0}$, $s$, $V_{0}$, $S$, and the
self-energies $\Sigma_{0}$ and $\Sigma$ are generally different
for protons and neutrons.

\subsection{Nuclear matter}

Solutions of the Dirac equation are given by the plane-wave states
\begin{equation}
 \psi(\vec{p},\sigma,\tau) = u(\vec{p},\sigma,\tau) 
 \exp \left( -i p_{\mu} x^{\mu} \right)
\end{equation}
for a nucleon with four momentum $p^{\mu} = (E,\vec{p})$,
energy $E>0$. The positive-energy four-spinor is denoted by
$u(\vec{p},\sigma,\tau)$ with spin and isospin quantum numbers
$\sigma$ and $\tau$, respectively. The energy $E_{\tau}$
of a proton ($\tau=1$) or neutron ($\tau = -1$) with momentum $p$
is found in the same way
as in Ref.\ \cite{Typ03} by replacing $X_{00}$ with $Y_{00}$,
$Y_{0}$ with $Z_{0}$, and setting $W=0$ in the corresponding expressions..
It is given by
\begin{eqnarray}
 E_{\tau} & = & \left(1+Y_{00}\right)^{-1}
 \\ \nonumber & & \times
 \left[ V_{0\tau} + \frac{1}{\sqrt{1-B^{2}}}
 \left( A_{\tau}B + \sqrt{A_{\tau}^{2}+p^{2}}\right)\right]
\end{eqnarray}
with the quantities
\begin{equation}
 A_{\tau} = \frac{m-S_{\tau}+BV_{0\tau}}{\sqrt{1-B^{2}}} 
\end{equation}
and
\begin{equation}
 B = \frac{Z_{0}}{1+Y_{00}} \: .
\end{equation}
The energy density $\varepsilon$ and the pressure $p$ are calculated
from the energy-momentum tensor. We find
\begin{eqnarray}
 \varepsilon & = & 
 \sum_{\tau} \left( \varrho^{E}_{\tau} - \varrho^{M}_{s\tau} \right)
   + m P_{s}
 \\ \nonumber & & 
 +\Gamma_{\omega} \omega_{0} J_{0} 
 +\Gamma_{\rho} \vec{\rho}_{0} \cdot \vec{J}_{0}
 -\Gamma_{\sigma} \sigma P_{s}
 -\Gamma_{\delta} \vec{\delta} \cdot \vec{P}_{s}
  \\ \nonumber & & 
 -\frac{1}{2} 
 \left[ m_{\omega}^{2} \omega_{0} ^{2}
  + m_{\rho}^{2}\vec{\rho}_{0}^{2} 
  - m_{\sigma}^{2} \sigma^{2} 
  - m_{\delta}^{2} \vec{\delta}^{2}
 \right]
\end{eqnarray}
and
\begin{eqnarray} \label{eq:pres}
 p & = &
\frac{1}{3} \sum_{\tau} \left(  \varrho^{E}_{\tau} 
 - \varrho^{M}_{s\tau} \right)
 +  \Sigma_{R0}j_{0} 
 \\ \nonumber & & 
 +\frac{1}{2} 
 \left[ m_{\omega}^{2} \omega_{0} ^{2}
  + m_{\rho}^{2}\vec{\rho}_{0}^{2} 
  - m_{\sigma}^{2} \sigma^{2} 
  - m_{\delta}^{2} \vec{\delta}^{2}
 \right] \: .
\end{eqnarray}
The densities $\varrho^{E}_{\tau}$ and $\varrho^{M}_{s\tau}$ are
calculated similarly as in Ref.\ \cite{Typ03}. 
They are given by
\begin{equation}
 \varrho^{E}_{\tau} = 
 \frac{2A_{\tau}BI^{\tau}_{1}+A_{\tau}^{2}B^{2}I^{\tau}_{2}
 +I^{\tau}_{3}}{\pi^{2}(1+Y_{00})(1-B^{2})^{3/2}}
\end{equation}
and
\begin{equation}
 \varrho^{M}_{s\tau} = 
 \frac{2A_{\tau}BI^{\tau}_{1}+A_{\tau}^{2}I^{\tau}_{2}
 +B^{2}I^{\tau}_{3}}{\pi^{2}(1+Y_{00})(1-B^{2})^{3/2}}
\end{equation}
with the integrals
\begin{eqnarray}
 I^{\tau}_{1} & = & \frac{1}{3} \left(p^{F}_{\tau}\right)^{3} \: ,
 \\
 I^{\tau}_{2} & = & \frac{1}{2} \left[p^{F}_{\tau}E^{F}_{\tau}
 - A_{\tau}^{2} \ln \frac{p^{F}_{\tau}+E^{F}_{\tau}}{A_{\tau}} \right] \: ,
 \\ 
 I^{\tau}_{3} & = & \frac{3}{4} E^{F}_{\tau} I^{\tau}_{1} 
 + \frac{1}{4} A_{\tau}^{2} I^{\tau}_{2} \: ,
\end{eqnarray}
c.f.\ Ref.\ \cite{Typ03}. The Fermi momentum $p^{F}_{\tau}$ 
is determined from $\varrho_{\tau} = I^{\tau}_{1}/\pi^{2}$
by the proton and neutron densities. The energy
$E^{F}_{\tau} = \sqrt{A_{\tau}^{2} + \left(p^{F}_{\tau}\right)^{2}}$
also depends on the quantity $A_{\tau}$.
The standard scalar and the vector densities can be expressed as
\begin{eqnarray}
 \varrho_{s\tau} & = & 
 \frac{BI^{\tau}_{1}+A_{\tau}I^{\tau}_{2}}{\pi^{2}(1+Y_{00})(1-B^{2})^{1/2}}
 \: ,
 \\
 j_{0\tau} & = & 
 \frac{I^{\tau}_{1}+A_{\tau}BI^{\tau}_{2}}{\pi^{2}(1+Y_{00})(1-B^{2})^{1/2}}
 \: .
\end{eqnarray}
We also need the kinetic densities
\begin{equation}
 t_{00}^{D} = \frac{\sum_{\tau}\varrho^{E}_{\tau}}{1+Y_{00}}
\end{equation}
and
\begin{equation}
 j_{0}^{D} = \frac{\sum_{\tau}\varrho^{E}_{s\tau}}{1+Y_{00}}
\end{equation}
with
\begin{equation}
 \varrho^{E}_{s\tau} = 
 \frac{A_{\tau}(1+B^{2})I^{\tau}_{1}+A_{\tau}^{2}BI^{\tau}_{2}
 +BI^{\tau}_{3}}{\pi^{2}(1+Y_{00})(1-B^{2})^{3/2}}
\end{equation}
in the field equations of the mesons.
In equation (\ref{eq:pres}) 
for the pressure $p$ a rearrangement contribution
appears explicitly. It guarantees that
the D${}^{3}$C model is thermodynamically
consistent and the Hugenholtz-van Hove theorem \cite{Hug58} 
holds as in the case of the DC and DD models.

\subsection{Finite nuclei}

The densities and fields in the ground state of finite nuclei 
do not vary with time but they depend on the spatial coordinates.
In the Hartree approximation the state of the nucleus is described
by a product of single-particle states $|\phi_{i}\rangle$ 
with single-particle energies
$\varepsilon_{i}$ (including the rest mass $m$) that are solutions of the 
Dirac equation. From the nucleon wave functions the various densities
are calculated by a summation over the contributions from the
individual nucleons. The meson fields are found by solving the 
corresponding field equations with space-dependent source terms
with appropriate boundary conditions. 

The energy of an atomic nucleus in the D${}^{3}$C model is given
by
\begin{eqnarray} \label{eq:E_nuc}
 E & = & 
 \sum_{i} w_{i} \varepsilon_{i}
 +  \frac{1}{2} \int d^{3}r \: \left[
   \left( \Gamma_{\sigma}+ 2 \Gamma^{\prime}_{\sigma} j_{0}\right) 
  \sigma P_{s} \right.
 \\ \nonumber & & 
  + \left( \Gamma_{\delta} + 2 \Gamma^{\prime}_{\delta} j_{0}\right)
 \vec{\delta} \cdot \vec{P}_{s}
  - \left( \Gamma_{\omega} + 2 \Gamma^{\prime}_{\omega} j_{0}\right)
 \omega_{0} J_{0}
   \\ \nonumber & & \left.
  - \left( \Gamma_{\rho} + 2 \Gamma^{\prime}_{\rho} j_{0}\right) 
 \vec{\rho}_{0} \cdot \vec{J}_{0}
  - \Gamma_{\gamma}  A_{0} J_{\gamma 0} \right]
 \\ \nonumber & &   
 +  \int d^{3}r \: \left[
    \left( t^{D}_{00} - j^{M^{\ast}}_{0} \right) 
 \left( 1 + \frac{\Gamma^{\prime}_{V}}{\Gamma_{V}}j_{0}\right) 
 Y_{00} \right.
 \\ \nonumber & &  \left.
  -  \left( j^{D}_{0}-\varrho^{M^{\ast}}_{s} \right)
 \left( 1 + \frac{\Gamma^{\prime}_{S}}{\Gamma_{S}}j_{0}\right) 
 Z_{0} \right]
\end{eqnarray}
with single-particle occupation numbers $w_{i}$
(in the present case without pairing $w_{i}=0$ or $1$, respectively).
Besides the usual rearrangement 
contributions due to the density dependence of the
nucleon-meson couplings $\Gamma_{\sigma}$, $\Gamma_{\delta}$,
$\Gamma_{\omega}$, and $\Gamma_{\rho}$, additional terms appear
with the fields $Y_{00}$ and $Z_{0}$ that contain
derivatives of the couplings $\Gamma_{V}$ and $\Gamma_{S}$.

Here, we consider only spherical nuclei. 
The self-consistent calculation of the spherical nuclei
was performed in coordinate space in an angular momentum basis
with a discretization of the radial coordinate and
a mesh spacing of $0.1$~fm
using similar procedures as in \cite{Typ99}.
A correction of the Coulomb field was introduced as in \cite{Typ99}.
Because the translational symmetry in the calculation is broken
a c.m.\ correction has to be added to the energy (\ref{eq:E_nuc}).
It has been microscopically calculated 
in the non-relativistic approximation
\begin{equation}
 E_{cm} = -  \frac{\langle \vec{P}^{2}\rangle}{2mA}
\end{equation}
with the c.m.\ momentum $\vec{P} = \sum_{i=1}^{A} \vec{p}_{i}$
from the single-particle wave functions \cite{Ben00}.
Nuclear radii are also corrected
for the c.m.\ motion. 
In case of the charge distribution, corrections
due to the c.m.\ motion and the formfactors of protons and neutrons
are considered in the calculation of the charge formfactor \cite{Rei86,Ruf88}.

\section{\label{sec:3} Parametrization}

In order to calculate actual properties of nucler matter and atomic 
nuclei the parameters entering the Lagrangian have to be specified.
For the mass of the nucleon, the $\omega$ and the $\rho$ meson the 
conventional values $m=939$~MeV, $m_{\omega} = 783$~MeV and 
$m_{\rho} = 763$~MeV were used. The mass of the $\sigma$ meson $m_{\sigma}$
is treated as a free parameter. The $\delta$ meson is neglected
in the present work.
Except for the electromagnetic
coupling constant $\Gamma_{\gamma}$, 
all parameters in the coupling functions $\Gamma_{i}$
and the mass of the $\sigma$ meson $m_{\sigma}$
were obtained from a fit to properties of the 
eight doubly 
magic spherical nuclei ${}^{16}$O, ${}^{24}$O, ${}^{40}$Ca, ${}^{48}$Ca,
${}^{56}$Ni, ${}^{100}$Sn, ${}^{132}$Sn, ${}^{208}$Pb
that include nuclei close to and far from the valley of stability.
Other nuclei were not considered because effects due to
nucleon pairing have to included explicitly in this case
adding more parameters in the fit.

There is no unique and generally accepted strategy for the fitting
procedure \cite{Bue04}.
In the present $\chi^{2}$ fit of the parameters
experimental data for
binding energies, charge radii, diffraction radii, surface
thicknesses, 
and spin-orbit splittings were taken into account \cite{Rei86,Ruf88}.
Furthermore, it was required to reproduce
the experimental value of $0.20(4)$~fm for the neutron skin thickness
in ${}^{208}$Pb, i.e.\ the difference $r_{n}-r_{p}$ of the neutron
and proton rms radii \cite{Sta94}. 
In total there are 32 experimental data points
used in the fit.

For a comparison of the D${}^{3}$C model with standard density-dependent
models, a new parametrization for DD models was derived under the same
conditions in the fit as the D${}^{3}$C model with the same
set of nuclei and experimental data. 
In this fit the nuclear incompressibility was fixed to
$K = 240$~MeV as in the original model in Ref.\ \cite{Typ99} where only
nuclear binding energies were considered.
A fit without this constraint leads to unacceptably large values of 
$K \approx 300$~MeV.

The D${}^{3}$C model with momentum-dependent self-energies contains
two additional coupling functions, $\Gamma_{V}$ and $\Gamma_{S}$,
as compared to usual DD models. They allow
to fit additional properties of the nuclear system under consideration.
Here, it was required that the optical potential (\ref{eq:pot_cen})
in symmetric nuclear matter at saturation density
assumes the value $50$~MeV at a nucleon energy of $1$~GeV, a value
that is typical for Dirac phenomenology \cite{Ham90,Coo93}.

The density dependence of the coupling functions is written as
\begin{equation}
 \Gamma_{i}(\varrho_{v}) = \Gamma_{i}(\varrho_{\rm ref}) f_{i}(x)
\end{equation}
with the coupling constants $\Gamma_{i}(\varrho_{\rm ref})$ at a reference
density $\varrho_{\rm ref}$ and suitable functions $f_{i}$ that
depend on the ratio
\begin{equation}
 x = \frac{\varrho_{v}}{\varrho_{\rm ref}}
\end{equation}
with the vector density $\varrho_{v}$. Note that the vector density
in the D${}^{3}$C model is not identical with the baryon density $\varrho$.
In the DD model the reference density is just the saturation density
$\varrho_{\rm sat}$ of symmetric nuclear matter.
The reference density in the D${}^{3}$C model
is different from $\varrho_{\rm sat}$
but it corresponds to the vector density $\varrho_{v}$ determined at
saturation of symmetric nuclear matter.

The functional form of $f_{i}(x)$
for the density dependence of the $\sigma$, $\omega$ and $\rho$
meson is assumed to be the same as introduced in
Ref.\ \cite{Typ99} and later used in the parametrizations
DD-ME1 \cite{Nik02}, DD-ME2 \cite{Lal05},
and PKDD \cite{Lon04}. 
For the $\sigma$ and $\omega$ meson it is given by
the rational function
\begin{equation} \label{eq:ddso}
 f_{i}(x) = a_{i} \frac{1+b_{i}(x+d_{i})^{2}}{1+c_{i}(x+d_{i})^{2}}
\end{equation}
with constants $a_{i}$, $b_{i}$, $c_{i}$, and $d_{i}$.
In order to reduce the number of free parameters it is required that
the functions $f_{\sigma}$ and $f_{\omega}$ obey the 
conditions $f_{\sigma}(1)=f_{\omega}(1)=1$, 
$f_{\sigma}^{\prime}(0)=f_{\omega}^{\prime}(0)=0$, and
$f_{\sigma}^{\prime\prime}(1)=f_{\omega}^{\prime\prime}(1)$.
For the $\rho$ meson a simple exponential law 
\begin{equation}
 f_{\rho}(x) = \exp\left[-a_{\rho}\left( x-1 \right) \right]
\end{equation}
is assumed with one free parameter $a_{\rho}$ as in Ref.\ \cite{Typ99}.

In the D${}^{3}$C model the quantities $Y_{00}$ and $Z_{0}$
are responsible for the energy dependence of the self-energies.
Assuming constant couplings $\Gamma_{V}$ and $\Gamma_{S}$
both $Y_{00}$ and $Z_{0}$ show an approximate $\varrho^{2}$ dependence at
low densities $\varrho$. This suggests to introduce 
the power law
\begin{equation} \label{eq:pl}
 f_{i}(x) = x^{-a_{i}}
\end{equation}
as the functional form in order to be able 
to arbitrarily adjust the density dependence of the energy dependence.
In this work the parameters $a_{V}$ and $a_{S}$ 
are set to unity because in this case
the second integral in the energy (\ref{eq:E_nuc}) of an atomic nucleus
does not contribute and the calculation is simplified.

In total there are ten free parameters in the D${}^{3}$C model
and eight in the DD model. Since several of the parameters are
strongly correlated \cite{Bue04} it is convenient to perform the fit 
not directly in all of these parameters directly
but to use saturation properties
of symmetric nuclear matter (see subsection \ref{subsec:41}) 
as independent variables in the fit and convert
these quantities to the coupling constants analytically. 
Considering the conditions for the 
neutron skin thickness in ${}^{208}$Pb and the optical potential,
the minimum of $\chi^{2}$ has to be found in an eight- (seven-)
dimensional parameter space.

\begin{table}
\caption{\label{tab:1}
Mass of the $\sigma$ meson $m_{\sigma}$,  
parameters of the coupling functions and
reference density $\varrho_{\rm ref}$ in the DD and D${}^{3}$C model.
}
\begin{ruledtabular}
\begin{tabular}{ccc}
 & DD & D${}^{3}$C \\
 \hline
 $m_{\sigma}$ [MeV] & 547.204590 & 556.986206 \\
 $\Gamma_{\sigma}(\varrho_{\rm ref})$ & 10.685257 & 11.027474 \\
 $a_{\sigma}$ & 1.371545 & 1.846341 \\
 $b_{\sigma}$ & 0.644063 & 3.520730 \\
 $c_{\sigma}$ & 1.034552 & 7.071799 \\
 $d_{\sigma}$ & 0.567627 & 0.217107\\
 $\Gamma_{\omega}(\varrho_{\rm ref})$ & 13.312280 & 13.750549 \\
 $a_{\omega}$ & 1.385567 & 2.004516 \\
 $b_{\omega}$ & 0.521724 & 4.563703 \\
 $c_{\omega}$ & 0.869983 & 9.864792 \\
 $d_{\omega}$ & 0.618991 & 0.183821\\
 $\Gamma_{\rho}(\varrho_{\rm ref})$ & 3.639023 & 3.917462 \\
 $a_{\rho}$ & 0.4987 & 0.4220 \\
 $\Gamma_{V}(\varrho_{\rm ref})$ & 0.0 & 302.188656 \\
 $\Gamma_{S}(\varrho_{\rm ref})$ & 0.0 & -21.632122 \\
 $\varrho_{\rm ref}$ [fm${}^{-3}$] & 0.148746 & 0.128941 \\
\end{tabular}
\end{ruledtabular}
\end{table}


\section{\label{sec:4} Results}

The actual values of the model parameters as determined in the fit are
given in Table \ref{tab:1}. The main difference between the DD and the
D${}^{3}$C model are the values of  the parameters $a_{i}$, $b_{i}$, $c_{i}$,
and $d_{i}$ for $i=\sigma, \omega$ in the functional form of the
density dependence (\ref{eq:ddso}). This corresponds to 
a much stronger increase of the $\sigma$ and $\omega$ coupling
for $\varrho \to 0$ in the D${}^{3}$C model as compared to
the DD parametrization. The density dependence of the $\rho$ meson, however,
is smaller for D${}^{3}$C  than for  DD. Obviously, the constants
$\Gamma_{V}(\varrho_{\rm ref})$ and $\Gamma_{S}(\varrho_{\rm ref})$
are non-vanishing only in the D${}^{3}$C model. Note that $\Gamma_{V}$
is more than ten times larger than $\Gamma_{S}$. 
Correspondingly, the energy dependence of the vector self-energy
is much larger than the energy dependence of the scalar self-energy
in the present parametrization of the D${}^{3}$C model.
The negative sign of $\Gamma_{S}(\varrho_{\rm ref})$ indicates that
the scalar self-energy rises weakly with the energy of the nucleon.
In contrast the vector self-energy decreases much more strongly.
From Dirac phenomenology
\cite{Ham90,Coo93}
one would expect that both the scalar and the vector self-energy
decrease at larger nucleon energies. However, by calculating
the momentum-dependent contributions to the self-energies from
the Fock terms considering one-pion exchange (that is not included
in conventional RMF models), an opposite energy dependence
of the self-energies is expected \cite{Mar00} with an increase of
the scalar self-energy and a decrease of the vector self-energy. 
The result of the D${}^{3}$C model lies between these two cases.

\begin{table}
\caption{\label{tab:2}
Contributions of nuclear properties with assumed uncertainty
in the $\chi^{2}$ fit of the model 
parameters and total $\chi^{2}$ in the DD and D${}^{3}$C model.
}
\begin{ruledtabular}
\begin{tabular}{cccc}
 property & uncertainty & DD & D${}^{3}$C \\
 \hline
 binding energies      & 0.2 MeV  & 225.2 & 168.4 \\
 charge radii          & 0.01 fm  &  44.9 &  32.4 \\
 diffraction radii     & 0.01 fm  &  74.7 &  53.3 \\
 surface thicknesses   & 0.005 fm &  15.0 &  23.0 \\
 spin-orbit splittings & 0.2 MeV  &  95.4 &  60.3 \\
\hline
 total & & 455.3 & 337.4\\
\end{tabular}
\end{ruledtabular}
\end{table}

In Table \ref{tab:2} the contributions to the total $\chi^{2}$
by the various nuclear properties
in the fit of the parameters are given for the
DD and D${}^{3}$C model, respectively. In both cases the error in the
binding energies contributes approximately one half to the total $\chi^{2}$.
The D${}^{3}$C model gives a better description of all properties
except for the surface thickness with a small increase of the 
corresponding partial $\chi^{2}$. The total $\chi^{2}$ 
of the D${}^{3}$C model is about  27 \% smaller than in the 
DD model. This represents a considerable improvement in the description
of the experimental data. More details for finite nuclei will be
discussed in subsection \ref{subsec:42}.

\subsection{\label{subsec:41} Nuclear matter}

\begin{figure}
\begin{center}
\includegraphics[width=85mm]{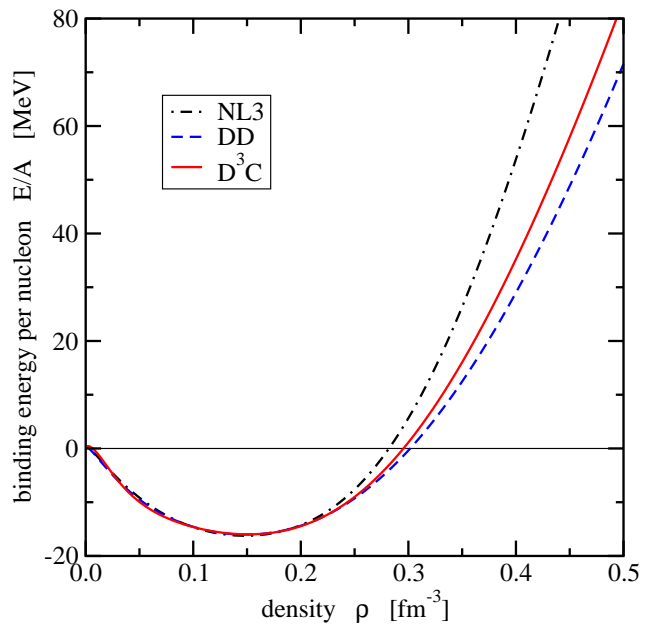}
\end{center}
\caption{\label{fig:eos_sym}
Equation of state for symmetric nuclear matter 
in various parametrizations.
}
\end{figure}

The equation of state (EOS) of symmetric nuclear matter, i.e.\ the binding 
energy per nucleon as a function of the density $\varrho$, is depicted in
Figure \ref{fig:eos_sym}. The EOS in the D${}^{3}$C model and the
DD model are compared to the result for the nonlinear NL3 parametrization
\cite{Lal97}
that is widely used in RMF calculations with considerable success.
All three curves are very similar below a density of approximately 
$0.2$~fm$^{-3}$. 
At high densities there is a noticable difference. 
The parametrization NL3 leads to the stiffest EOS
and the DD model has the softest EOS
with the D${}^{3}$C curve lying in between.

A more quantitative comparison of the models is provided by examining
the characteristic constants
in the expansion of the binding energy per nucleon
\begin{eqnarray}
 \frac{E}{A} & = &
 a_{V} + \frac{K_{\infty}}{18} \epsilon^{2} 
 - \frac{K^{\prime}}{162} \epsilon^{3} + \dots 
 \\ \nonumber & & 
 + \delta^{2} \left( J + \frac{L}{3} \epsilon
 + \frac{K_{\rm sym}}{18} \epsilon^{2} + \dots \right) 
\end{eqnarray}
in nuclear matter near saturation
\cite{Mye85,Nay90}.
The deviation of the density from saturation is quantified by
\begin{equation}
 \epsilon = \frac{\varrho-\varrho_{\rm sat}}{\varrho_{\rm sat}}
\end{equation}
and the neutron-proton asymmetry is given by
\begin{equation}
 \delta = \frac{\varrho_{n}-\varrho_{n}}{\varrho_{n}+\varrho_{p}}
 \: .
\end{equation}
In symmetric nuclear matter, i.e.\ $\delta = 0$, 
the binding energy at saturation $a_{V}$,
the incompressibility $K_{\infty}$, and the derivative $K^{\prime}$ of the
incompressibility determine the form of the EOS. 
For asymmetric nuclear matter, also the symmetry energy $J$, 
the derivative $L$ of the symmetry energy, and the symmetry
incompressibility $K_{\rm sym}$ are important. The values
of these quantities for the three models NL3, DD, and D${}^{3}$C are
given in Table {\ref{tab:3}. A detailed comparison shows that there are major 
differences between the parametrizations. The D${}^{3}$C model
has the softest EOS near saturation ($K_{\infty}$) but at higher densities
is becomes stiffer than the DD model because of the large negative 
$K^{\prime}$.
The saturation density is larger than in the DD and NL3 models where the
latter model also has a rather strong binding.

\begin{table}
\caption{\label{tab:3}
Properties of symmetric nuclear matter at saturation in various models.
}
\begin{ruledtabular}
\begin{tabular}{cccc}
 & NL3 & DD & D${}^{3}$C \\
 \hline
 $\varrho_{\rm sat}$ [fm${}^{-3}$] &  0.1482 & 0.1487 & 0.1510 \\
 $a_{V}$ [MeV]                     & -16.240 & -16.021 & -15.981\\
 $K_{\infty}$ [MeV]                &  271.5  & 240.0 & 232.5 \\
 $K^{\prime}$ [MeV]                & -203.0  & -134.6 & -716.8 \\
 $J$ [MeV]                         &  37.4   & 31.6 & 31.9 \\
 $L$ [MeV]                         &  118.5  & 56.0 & 59.3 \\
 $K_{\rm sym}$ [MeV]               &  100.9  & -95.3 & -74.7 \\
 $m^{\ast}/m$                      &  0.596  & 0.565 & 0.541 \\
 $m_{\rm eff}/m$                   &  0.655  & 0.628 & 0.710 \\
 $V_{\rm cen}$(1 GeV) [MeV]        &  282.2  & 310.0 & 50.0 \\
\end{tabular}
\end{ruledtabular}
\end{table}

\begin{figure}
\begin{center}
\includegraphics[width=85mm]{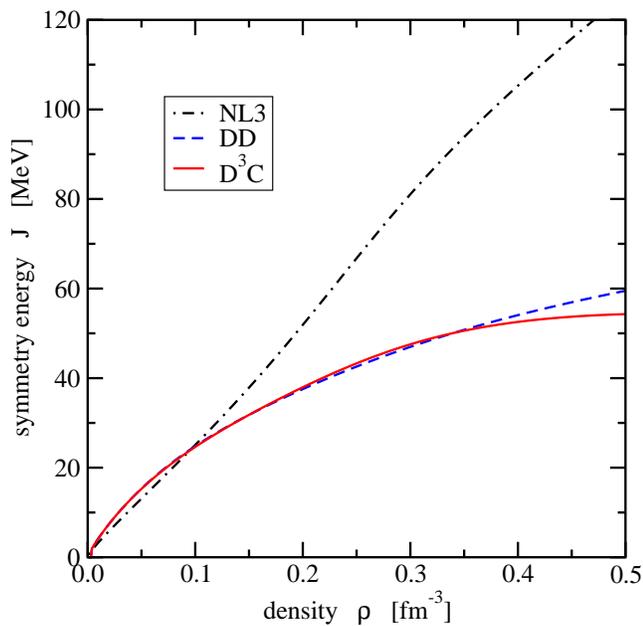}
\end{center}
\caption{\label{fig:J}
Symmetry energy $J$ in symmetric nuclear matter 
as a function of the nucleon density $\varrho$
in various parametrizations.
}
\end{figure}

The symmetry energy $J$ and
the derivative $L$ are considerably smaller in the D${}^{3}$C and the
DD model than in the NL3 parametrization. This is due to the
density dependence of the $\rho$ meson coupling. The actual values
are determined by the fit of  the 
neutron skin thickness in ${}^{208}$Pb. The differences in $L$ 
corresponds to a rather strong deviation
in the density dependence of the symmetry energy that is shown
in Figure \ref{fig:J}. In the NL3 model the symmetry energy rises almost
linearly with the density. In contrast, the DD and D${}^{3}$C model
exhibit a considerable flattening. This behavior 
is also reflected in the EOS for neutron matter as
presented in Figure \ref{fig:eos_nm}. Here, the NL3 model shows
a very stiff EOS. The shape of the neutron EOS in this model also
differs at low densities from the results in the DD and D${}^{3}$C model
that display very similar results for neutron matter.
It is worth noting that the symmetry energy and the neutron EOS 
in the three models are rather similar at a density near $0.1$~fm${}^{-3}$.
This value is approximately the neutron density in heavy atomic nuclei.
It would be rewarding to study the effects of the different 
equations of state 
on the properties of neutron stars but this is beyond the scope 
of the present work.

\begin{figure}
\begin{center}
\includegraphics[width=85mm]{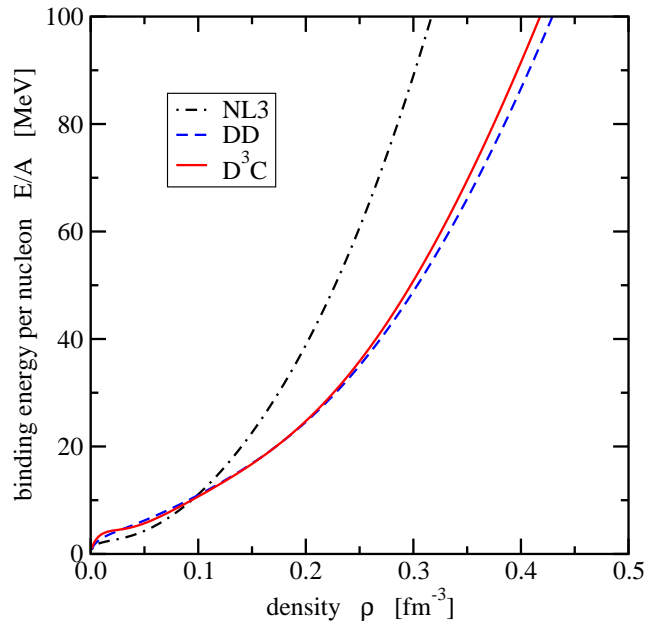}
\end{center}
\caption{\label{fig:eos_nm}
Equation of state for neutron matter 
in various parametrizations.
}
\end{figure}

Besides the parameters characterizing the EOS, also the Dirac
mass $m^{\ast}$ and the Landau mass $m_{\rm eff}$ 
(at the Fermi momentum) are of
great interest. The values of these quantities 
for saturated symmetric nuclear matter are also given
in Table \ref{tab:3}. The D${}^{3}$C model has the smallest 
value for $m^{\ast}$ even below that of the DD model but the
Landau mass of $0.710$ nucleon masses is the largest of the three
models. 
The $\chi^{2}$ of the fit is actually not very sensitive to a change in the
Landau mass. In principle it is possible to adjust $m_{\rm eff}$
to even larger values
at the Fermi surface in saturated symmetric nuclear matter
in the D${}^{3}$C model.

The density dependence of $m^{\ast}$ and $m_{\rm eff}$ in
symmetric nuclear matter is shown in Figure \ref{fig:meff}. 
The relativistic effective mass drops strongly with increasing
density. At low densities the D${}^{3}$C model is similar to the
DD model whereas at high densities it follows more closely the
NL3 model. At low densities the Landau mass decreases
like the relativistic effective mass with increasing density.
However, at higher densities an increase is observed that is easily
explained considering the dependence on the Fermi momentum 
in equations (\ref{eq:meff}) and (\ref{eq:meff2}). 
In the D${}^{3}$C model this increase
is very pronounced and the absolute values of $m_{\rm eff}$ are 
much larger than in the other parametrizations that are representative
for standard RMF models. Correspondingly,
the density of states at the Fermi surface in the D${}^{3}$C model
is considerably higher than that in other RMF models.
From the experience in the fitting of the parameters it is found that
the requirement of a larger Landau mass at saturation than in the present
D${}^{3}$C model leads to an even softer equation of state for symmetric
nuclear matter if one also demands an optical potential of $50$~MeV
at 1~GeV kinetic nucleon energy. Thus, this last condition should be
relaxed in a more extensive parameter fit.

\begin{figure}
\begin{center}
\includegraphics[width=85mm]{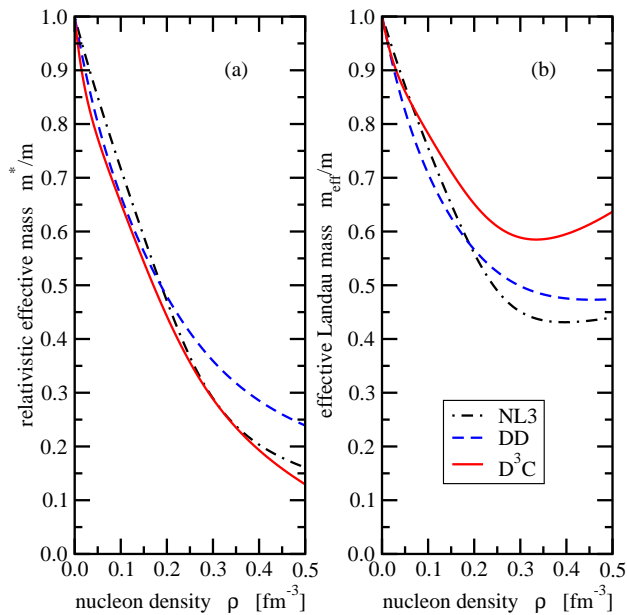}
\end{center}
\caption{\label{fig:meff}
Relativistic effective mass $m^{\ast}$ (a) and effective
Landau mass $m_{\rm eff}$ at the Fermi momentum (b) 
in units of the nucleon mass
$m$ in symmetric nuclear matter as a function of the density
in various parametrizations.
}
\end{figure}

Another major advantage of the D${}^{3}$C model is the possibility for a
very reasonable description of the Schr\"{o}dinger-equivalent
central opical potential as depicted in Figure \ref{fig:vopt}. 
At low kinetic energies of the nucleon it rises nearly linearly
as the empirical optical potential of nuclear matter extracted from Dirac
phenomenology for elastic proton-nucleus scattering  \cite{Coo93,Ham90}.
At higher nucleon energies it shows a similar saturation with
reasonable absolute values. In standard
RMF models like NL3 and DD without momentum-dependent self-energies the optical
potential rises linearly with the nucleon energy. It approaches 
unrealistically high values 
around 300~MeV at 1~GeV kinetic energy (see also Table \ref{tab:3}).
Because the optical potential
in the D${}^{3}$C model is a quadratic function of the nucleon
energy, it will ultimately decrease at higher energies und 
become unrealistic. However, this behavior is not relevant for 
most applications, e.g.\ finite nuclei or high-density nuclear matter,
because the relevant Fermi momenta and corresponding nucleon energies
are still small enough in these cases.

\begin{figure}
\begin{center}
\includegraphics[width=85mm]{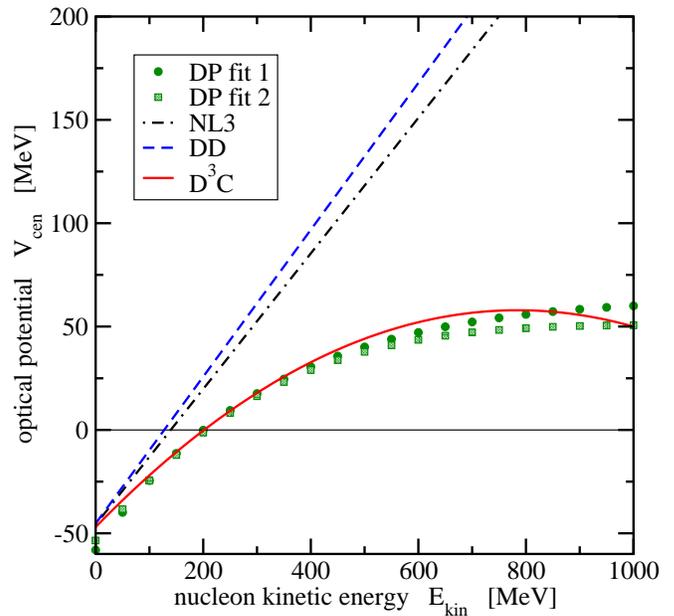}
\end{center}
\caption{\label{fig:vopt}
Schr\"{o}dinger-equivalent 
central optical potential $V_{\rm cen}$ in symmetric nuclear
matter as a function of the nucleon kinetic energy $E_{\rm kin} = E-m$
in various parametrizations and from Dirac phenomenology (DP) of elastic
proton-nucleus scattering \protect\cite{Coo93,Ham90}.
}
\end{figure}

\subsection{\label{subsec:42} Finite nuclei}

\begin{table}
\caption{\label{tab:4}
Total binding energies (in MeV) of the eight fit nuclei 
in the experiment \protect\cite{Aud03}
and in the models DD and D${}^{3}$C.
}
\begin{ruledtabular}
\begin{tabular}{cccc}
 Nucleus & Exp. & DD & D${}^{3}$C \\
 \hline
 ${}^{16}$O   &  -127.619 &  -128.064 &  -127.040 \\
 ${}^{24}$O   &  -168.382 &  -169.114 &  -169.178 \\
 ${}^{40}$Ca  &  -342.052 &  -342.505 &  -342.581 \\
 ${}^{48}$Ca  &  -415.990 &  -414.876 &  -415.047 \\
 ${}^{56}$Ni  &  -483.992 &  -481.924 &  -482.486 \\
 ${}^{100}$Sn &  -824.794 &  -826.255 &  -826.079 \\
 ${}^{132}$Sn & -1102.851 & -1103.359 & -1103.478 \\
 ${}^{208}$Pb & -1636.430 & -1636.030 & -1635.893 \\
\end{tabular}
\end{ruledtabular}
\end{table}

From Table \ref{tab:2} it was already seen that the D${}^{3}$C model
improves the description of various properties of finite nuclei as
compared to the DD model demanding the same conditions in the fit.
In Table \ref{tab:4} the experimental binding energies 
\cite{Aud03} of the nuclei that were
considered in the fit are compared with the results of the DD and the
D${}^{3}$C model. In general, there is a good reproduction of the
experimental data. The absolute root-mean square deviation from the
experiment is $1.06$~MeV and $0.92$~MeV, respectively. These values 
are still larger than corresponding numbers from dedicated fits to masses
with rms deviations in the 
order of a few hundred keV, see, e.g.,  \cite{Gor01,Sam02,Lun03}. For a fair
comparison one has to bear in mind that the 
RMF parametrizations in this work are not just mass models.
They are constructed in order 
to describe a large number of different properties of nuclei and nuclear
matter.
Furthermore, a valid comparison is only possible when a larger set
of (also deformed) nuclei is taken into account. However, this
requires a correction for the rotational energy and 
a reasonable description of pairing effects that is left for future
investigations. 

The characteristic parameters of the nuclear shape and of the
charge form factor, respectively, i.e.\
the charge radius, the diffraction radius, and the surface thickness
\cite{Rei86,Ruf88}, are compared to experimental data 
\cite{Fri95,Nad94} in Figure \ref{fig:radii}.
Both the DD model and the D$^{3}$C model agree very well with the
available experimental data. At the same time the difference 
between the models is very small. They show the same trend 
for the radii and for the surface thickness. The largest deviation from
experiment is found for the surface thickness of ${}^{16}$O, a light nucleus
that is not expected to be described very well in a mean-field model.

\begin{figure}
\begin{center}
\includegraphics[width=85mm]{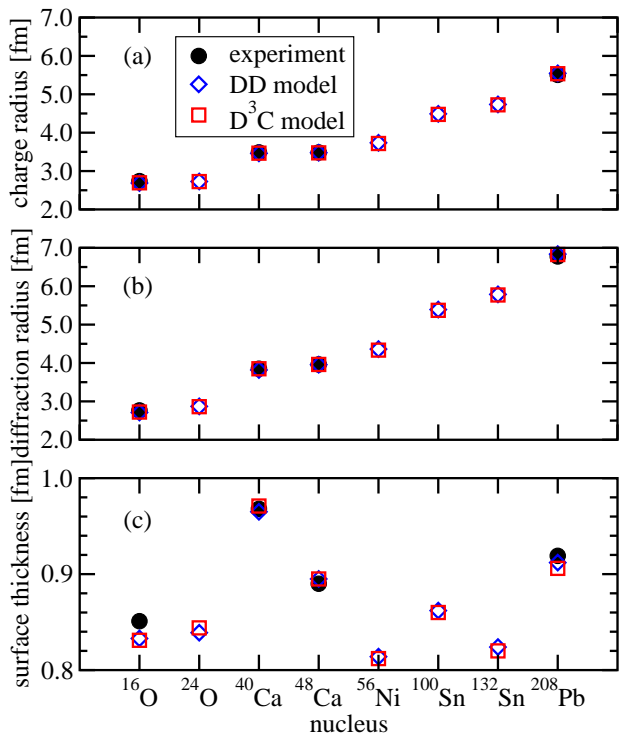}
\end{center}
\caption{\label{fig:radii}
Charge radius (a), diffraction radius (b), and
surface thickness (c) of the fit nuclei 
 in the DD model (blue diamonds) and in the D${}^{3}$C model
(red squares) compared with data from the experiment
(solid black dots) \protect\cite{Fri95,Nad94}.
}
\end{figure}

The last quantity that was considered in the fit is the spin-orbit
splitting of proton and neutron single-particle energies of
levels with the same principal quantum number and the same orbital
angular momentum. In Table \ref{tab:5} the results in the DD model
and the D${}^{3}$C model are compared to experimental data for levels
close to the Fermi surface. It is found that the spin-orbit splittings
of the D${}^{3}$C model improves the description in most cases when
compared to the DD model. This is possible because the relativistic
effective mass $m^{\ast}$ in the former model is even smaller than
in the latter model. At the same time, the Landau mass $m_{\rm eff}$ 
in the D${}^{3}$C model is considerably larger than in the DD model.
Correspondingly, the level density of the D${}^{3}$C model is higher.
This compression of the spectrum is easily seen when the distribution 
of single-particle levels
is compared for the three heaviest nuclei
in the fit  in Figure \ref{fig:spe}. 
Levels below the Fermi energy are shifted to higher 
energies and levels above the Fermi energy become more strongly bound.

\begin{table}
\caption{\label{tab:5}
Spin-orbit splitting (in MeV) of neutron ($\nu$) and proton
($\pi$) levels in the experiment
and in the models DD and D${}^{3}$C.
}
\begin{ruledtabular}
\begin{tabular}{ccccc}
 Nucleus & State & Exp. & DD & D${}^{3}$C \\
 \hline
 ${}^{16}$O   & $\nu$0p & 6.18 & 6.761 & 6.278 \\
              & $\pi$0p & 6.32 & 6.695 & 6.218 \\ \hline
 ${}^{48}$Ca  & $\nu$0f & 8.39 & 7.961 & 7.623 \\
              & $\nu$1p & 2.03 & 1.512 & 1.585 \\ \hline
 ${}^{56}$Ni  & $\nu$0f & 7.16 & 8.568 & 8.369 \\
              & $\nu$1p & 1.12 & 1.445 & 1.335 \\ \hline
 ${}^{132}$Sn & $\pi$1d & 1.74 & 1.996 & 1.786 \\
              & $\pi$0g & 6.09 & 6.563 & 6.171 \\ \hline
 ${}^{208}$Pb & $\nu$2p & 0.90 & 0.917 & 0.879 \\
              & $\pi$1d & 1.33 & 1.834 & 1.615 \\
              & $\pi$0h & 5.56 & 6.063 & 5.642 \\
\end{tabular}
\end{ruledtabular}
\end{table}

\begin{figure}
\begin{center}
\includegraphics[width=85mm]{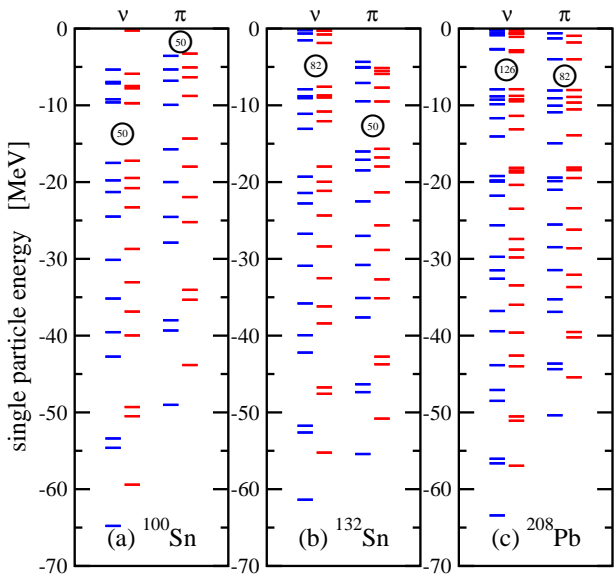}
\end{center}
\caption{\label{fig:spe}
Neutron ($\nu$) and proton ($\pi$) single particle energies 
for (a) ${}^{100}$Sn, (b) ${}^{132}$Sn, and (c) ${}^{208}$Pb in
the DD (blue lines, left) and D${}^{3}$C (red lines, right) model.
The position of the shell closure is denoted by open circles.
}
\end{figure}

\section{\label{sec:5} Conclusions and outlook}

In this work the 
conventional relativistic mean-field model with density-dependent
nucleon-meson couplings was extended by introducing a particular form
of couplings between the isoscalar meson fields and derivatives of
the nucleon fields. This approach leads to a linear momentum
dependence of the scalar and vector self-energies in the Dirac equation
for the nucleon. The parameters of the model were determined by a fit
to properties of finite nuclei. It was possible to
improve the description of binding
energies, nuclear shapes and spin-orbit splittings of single-particle
levels. 
The characteristic parameters of nuclear matter in the D${}^{3}$C model
are shifted closer to the values of non-relativistic Skyrme
Hartree-Fock models \cite{Ben03}.
Most noticable was the increase of the effective Landau
mass and, correspondingly, the level density at the Fermi surface
as compared to standard RMF models. At the same time, the momentum dependence
of the self-energies leads to a Schr\"{o}dinger-equivalent optical 
potential in symmetric nuclear matter that follows closely the empirical data
whereas standard RMF models fail.

The introduction of couplings of the meson fields to derivaties of the 
nucleon field is a purely phenomenological approach. The goal 
was to investige 
possible extensions of the RMF model and their effect on various 
quantities in nuclear matter and finite nuclei. The origin or the underlying
mechanism of the additional couplings is not really relevant in the present 
context. They are seen only as an effective way to introduce some 
desired effects in the spirit of density-functional theory. This approach 
can be compared with the medium dependence that can be parametrized in 
relativistic mean field models by self-interactions of the meson fields
in non-linear models, by density dependent couplings, or even by density 
dependent meson masses. 
These approaches are more or less equivalent 
when the final results for properties of nuclei are compared, however
the underlying mechanisms are quite different.

In principle, of course, one could go beyond the mean field model
and discuss in a systematic diagrammatic expansion modifications of
the self energies. But this 
was not the topic of the work.
In the present 
paper, I 
wanted to stay on a purely phenemenological level 
in a simple self-consistent approach that has all the necessary features 
that make it possible to apply the model in practical calculations 
with reasonable effort to nuclear matter and finite nuclei. The results 
show that the approach works surprisingly well. The modification of the 
Lagrangian density can be discussed and treated without reference to 
more fundamental mechanisms that could generate the corresponding terms. 

The density dependence of the momentum dependence is determined
in the present model by the choice of the parameters $a_{V}$ and
$a_{S}$ in Eq.\ (\ref{eq:pl}). Other values than $a_{V}=a_{S}=1$
as in the present model should be considered.
A fit of the model parameters
by fixing the Landau mass at the Fermi momentum but varying the
absolute value of the Schr\"{o}dinger-equivalent optical
potential is conceivable.
The parameter space can be explored more thoroughly allowing for
a simultaneous variation of the optical potential and of the Landau mass. 
The study of deformed nuclei is a further test of the model.
However, it will require a larger computational effort and the
consideration of pairing effects. It will
give more insight into the quality of the description. 
Applications to neutron stars and to simulations of heavy-ion collisions 
are also possible.
Extensions of the model to introduce an additional isospin dependence of 
the momentum dependence are obvious. 

\acknowledgments

This paper profited by the presentations and discussions at the
Workshop on relativistic density functional theory for nuclear
structure at the Institute for Nuclear Theory in Seattle, Washington,
September 20 - 24, 2004. The author acknowledges the discussion
with the participants.


\end{document}